\documentstyle[preprint,aps]{revtex}
\begin{document}
\title{Influence of nearly resonant light on the scattering length
     in low-temperature atomic gases}
\author{P.O. Fedichev$^{1,2}$, Yu. Kagan$^{1,2}$, G.V. Shlyapnikov$^{1,2}$
and J.T.M. Walraven$^{1,3}$}
\address{
{\em (1)} {Van der Waals - Zeeman Institute, University of Amsterdam,\\
Valckenierstraat 65-67, 1018 XE Amsterdam, The Netherlands}\\
{\em (2)} { Russian Research Center Kurchatov Institute,\\
Kurchatov Square, 123182 Moscow, Russia}\\
{\em (3)} {\ Ecole Normale Sup\'{e}rieure, Paris, 75231 France}
}
\date{\today }
\maketitle
\begin{abstract}
We develop the idea of manipulating the scattering length $a$ in
low-temperature atomic gases by using nearly resonant light.
As found, if the incident light is close to resonance with one of
the bound $p$ levels of electronically excited molecule, then
virtual radiative transitions of
a pair of interacting atoms to this level can significantly change
the value and even reverse the sign of $a$.
The decay of the gas due to photon recoil, resulting from the
scattering of light by single atoms,
and due to photoassociation can be minimized by selecting the
frequency detuning and the Rabi frequency.
Our calculations show the feasibility of optical manipulations
of trapped Bose condensates through a light-induced change in the
mean field interaction between atoms, which is illustrated for $^7$Li.

\end{abstract}


The recent successful experiments on Bose-Einstein condensation (BEC) in
magnetically trapped gases of Rb \cite{Cor95}, Li \cite{Hul95} and
Na \cite{Ket95} have generated a lot of interest in macroscopic quantum
behavior of atomic gases at ultra-low temperatures.
These experiments were enabled by
efficient evaporative \cite{Ket96} and optical
cooling \cite{Coh95} methods combined to reach the necessary
temperatures ($T\lesssim 1~\mu $K) and densities $10^{12}\lesssim n\lesssim
10^{14}$ cm$^{-3}$.

A principal question for BEC in atomic gases concerns the sign of
the scattering length $a$ for the pair elastic interaction.
For $a>0$ elastic interaction between atoms is repulsive and
the Bose condensate is stable with respect to this
interaction.
If $a<0$ elastic interaction is attractive and this is the
origin of a collapse of the condensate in a homogeneous gas\cite{LL}.
For trapped gases with $a<0$ the situation is likely to be the same,
provided the interaction between particles exceeds the level spacing
in the trapping field \cite{Ruprecht,Kagan95}.
If this interaction is much smaller than the level spacing,
there is a gap for one-particle excitations and it
is possible to form a metastable Bose-condensed state \cite{Kagan95}.
Among the alkalis there are atomic gases with positive as well as with
negative $a$ \cite{Verhaar}. Also the magnetic field dependence of $a$
was predicted \cite{VerhaarB}.

In this Letter we develop the idea of manipulating the value and the
sign of the scattering length by using nearly resonant light.
Since changing $a$
directly affects the mean field interaction between the atoms, this offers a
possibility to investigate macroscopic quantum phenomena associated with
BEC by observing the evolution of a Bose condensed gas in response to light.
Also optical control of cold elastic collisions is attracting
interest \cite{Bag96,Jul96}.
The physical picture of the influence of the light field on the elastic
interaction between atoms is the following:
A pair of atoms absorbs
a photon and undergoes a virtual transition to an electronically excited
quasimolecular state.
Then it reemits the photon and returns
to the initial electronic state at the same kinetic energy.
As the interaction between
atoms in the excited state is much stronger than in the
ground state, already at moderate light intensities the scattering amplitude
can be significantly changed.

The presence of the light field also induces
inelastic processes, such as photon recoil and light
absorption in pair collisions (with regard to cold
collisions see refs. \cite{Jul93,Heinzen} for review).
Photon recoil is the result of the scattering of light by single atoms.
At subrecoil temperatures, typical for achieving BEC, recoiling atoms
are lost as they overcome the confining barrier and escape from the trap.
The probability of light scattering by single atoms is proportional to
$(\Omega/\delta)^2$,
where $\Omega $ is the Rabi frequency and $\delta $ is the
frequency detuning of the light with respect to a single atom at rest. To
suppress the recoil losses the ratio $\Omega /\delta $ needs to be
sufficiently small. Then, for positive $\delta $, where the light is at
resonance with continuum states of excited quasimolecules, the change of $a$
will also be proportional to $(\Omega /\delta )^2$ and thus very small. To
have small recoil losses in combination with a significant change of $a,$
the detuning $\delta $ should be large and negative and not too far from a
vibrational resonance with one of the bound $p$ states of the
electronically excited molecule. However, the vicinity of the resonance
will also lead to photoassociation in pair collisions, followed
by spontaneous emission and loss from the trap.
Hence, the frequency detuning $\delta_{\nu}$
with respect to the $\nu$-th (nearest) vibrational resonance should greatly
exceed the line width of the resonance. We established that
it is possible to change the scattering length
substantially and even switch its sign without excessive recoil or
photoassociation losses. This will be illustrated for $^7$Li.

We consider low gas densities satisfying the condition
\begin{equation}            \label{n}
n\lambdabar^3\ll 1,
\end{equation}
where $2\pi\lambdabar$ is the optical wavelength. Then collective optical
effects \cite{Svistunov90} are absent, and at sufficiently low
temperatures the line broadening of optical transitions is determined by
the natural line width
$\Gamma=4d^2/3\hbar\lambdabar^3$, where $d$ is the transition dipole
moment. We analyze the influence of
incident light with large ($|\delta|\gg \Gamma$) and negative
frequency detuning on the interaction in a pair of atoms, with
vanishing wavevector of relative motion, ${\bf k}\rightarrow 0$.
The light frequency is assumed to be nearly
resonant with a highly excited vibrational $p$ level (with vibrational
quantum number $\nu$ and binding energy $\varepsilon _{\nu}$) in the
interaction potential
$V(r)$ of the attractive excited electronic state of the
quasimolecule, i.e., the frequency detuning with respect to this level,
$\delta_{\nu}=\delta -\varepsilon _{\nu}$, is much smaller than
the local vibrational level spacing $\Delta \varepsilon _{\nu}=\varepsilon
_{\nu}-\varepsilon _{\nu+1}$ (hereafter all frequencies are given in
energy units).
Then
radiative transitions of the pair from the ground electronic state to
the excited level $\nu$ are most important.
These transitions predominantly
occur at interparticle distances $r$ in the vicinity of the
outer turning point $r_t$ for the relative motion of atoms in the bound
state $v$, i.e., $V(r_t)=-\varepsilon_{\nu}$.
Unless $\varepsilon_{\nu}$ is very large, $r_t$ is determined by the
long-range part of $V(r)$, represented by the resonance dipole term.
If $\varepsilon_{\nu}$ and $|\delta|$ are still much larger than the
Zeeman and fine structure splitting, then at interparticle distances
relevant for radiative transitions the polarization
vector of the attractive excited quasimolecular state, ${\bf e}_\lambda $,
is parallel to the internuclear axis, and $V(r)=-2d^2/r^3$.
Hence, as $\varepsilon_{\nu}\sim |\delta|\gg \Gamma$,
we have $r_t\ll \lambdabar$.

For sufficiently large $\varepsilon_{\nu}$ and $\delta$
spontaneous emission of excited molecules predominantly
produces non-trapped atoms with kinetic energies of order
$\varepsilon_{\nu}$. These atoms practically do not interact with the
driving light and escape from the trap.
Therefore, the problem of finding the scattering length in the
presence of light is equivalent to a scattering problem which can be
be described in terms of wavefunctions of the ground and excited
electronic quasimolecular states. These states
are coupled by light, and spontaneous emission from the excited state
can be taken into account by adding the ``absorptive part''
$-i\Gamma$ (the spontaneous emission rate for molecules is twice
as large as that for single atoms)
to the interaction potential $V(r)$.

In the Born-Oppenheimer approximation the total wavefunction of the
quasimolecule in the presence of light can be written as $\phi({\bf
r})|g\rangle +\psi({\bf r})|e\rangle$, where $|g>$ and $|e>$ are the
electron wavefunctions of the ground and excited electronic states.
The wavefunctions of the relative motion of atoms in
these states, $\phi({\bf r})$ and
$\psi({\bf r})$, can be found from the system of coupled
Schr\"odinger equations:
\begin{eqnarray}
-\frac{\hbar^2}{m}\bigtriangleup_{{\bf r}}\phi({\bf r})+U(r)\phi({\bf
r})+\Omega\xi ({\bf r})\psi({\bf r})=0,
\label{phi} \\
-\frac{\hbar^2}{m}\!\!\bigtriangleup_{{\bf r}}\!\psi({\bf r})\!+\!(\!V(r)
\!-\!\!i\Gamma\!\!-\!\delta\!)\psi({\bf
r})\!+\!\!\Omega\xi ({\bf r})\phi({\bf r})\!=\!
0,\!\!\!  \label{psi}
\end{eqnarray}
where $\xi({\bf r})=({\bf e}_{\alpha}{\bf e}_{\lambda}({\bf r}))$,
$U(r)$ is the interaction potential in the
ground electronic state, and ${\bf e}_{\alpha}$ the polarization vector
of light. The Rabi frequency is defined as $\Omega=dE/\sqrt{2}$,
where $E$ is the amplitude of the electric field of light.
In Eqs. (\ref{phi}) and (\ref{psi}) we neglect the light shifts at infinite
separation between atoms and omit the recoil. These equations lead to the
integral equation for $\phi({\bf r})$:
\begin{equation}     \label{phiground}
\!\phi(\!{\bf r})\!\!=\!\phi_0(\!{\bf r})\!+\!\!\Omega^2
\!\!\!\!\int \!\!d{\bf r}''\!d{\bf r}'
G(\!{\bf r}''\!\!,\!{\bf r})\xi (\!{\bf r}'')\tilde G(\!{\bf r}''
\!\!,\!{\bf r}')\xi (\!{\bf r}')\phi(\!{\bf r}').\!\!
\end{equation}
Here $G({\bf r},{\bf r}')$ and $\tilde G({\bf r},{\bf r}')$ are the
Green functions of Eqs. (\ref{phi}) and (\ref{psi}) with $\Omega=0$.
The wavefunction  $\phi_0$ describes the relative motion of atoms
with zero energy for the potential $U(r)$ in the absence of light.
This function is a solution of Eq.(\ref{phi}) with $\Omega=0$.
The Green function $G({\bf r},{\bf r}')$ has the form
\begin{equation}         \label{G}
G({\bf r},{\bf r}')=\frac{m}{4\pi\hbar^2}
\times\left\{
\begin{array}{ll}
\phi_0(r)\tilde{\phi}_0(r'), & \mbox{$r<r'$} \\
\phi_0(r')\tilde{\phi}_0(r), & \mbox{$r>r'$}
\end{array}
\right.
\end{equation}
where $\tilde \phi_0(r)$ is a solution of the same
Schr\"odinger equation as that for $\phi_0(r)$, but contains only an
outgoing spherical wave at large $r$: $\tilde \phi_0(r)\rightarrow 1/r$
for $r\rightarrow\infty$.
As the frequency detuning of light was chosen such that
$|\delta_{\nu}|\ll \Delta\varepsilon_{\nu}$, the bound state $\nu$
should give the
dominant contribution to $\tilde G({\bf r},{\bf r}')$ and we may use
\begin{equation}           \label{Gex}
\tilde G({\bf r},{\bf r}')=-\psi_{\nu}({\bf r})
\psi_{\nu}^{*}({\bf r}')/(\delta_{\nu}+i\Gamma),
\end{equation}
where $\psi_{\nu}({\bf r})$ is the wavefunction of this state in the absence
of light.
Accordingly, the dependence of the rhs of Eq.(\ref{phiground}) on
$\phi({\bf r})$ will be only contained in the integral
$I=\int d^3{\bf r}'\phi({\bf r}')\xi({\bf r}')\psi_{\nu}^{*}({\bf r}')$.
Multiplying both sides of Eq.(\ref{phiground}) by $\xi({\bf
r})\psi_{\nu}^{*}({\bf r})$ and integrating over $d^3{\bf r}$, we express
$I$ through the overlap integral $I_0=\int d^3{\bf r}'\phi_0({\bf r}')
\xi({\bf r}')\psi_{\nu}^{*}({\bf r}')$. Then the exact solution of
Eq.(\ref{phiground}) is straightforward:
\begin{equation}              \label{phires}
\phi({\bf r})\!=\!\phi_0({\bf
r})\!-\!\frac{\Omega^2I_0\!\int \!\!d{\bf r}'\psi_{\nu}({\bf r}')
\xi({\bf r}')G({\bf r},{\bf r}')}{\delta_{\nu}+
(\Omega^2/\Delta\varepsilon_{\nu})\beta+i\Gamma}.
\end{equation}
The quantity $(\Omega^2/\Delta\varepsilon_{\nu})\beta$
describes the light-induced shift of the $\nu$-th vibrational
resonance, and the numerical factor
$\beta\!=\!\Delta\varepsilon_{\nu}\!\!\int \!\!d{\bf r}d{\bf r}'
G(\!{\bf r},\!{\bf r}')\xi({\bf r})
\psi_{\nu}^{*}({\bf r})\xi({\bf r}')
\psi_{\nu}({\bf r}')$.
As in the limit of zero energies only the $s$-wave contribution to
$\phi({\bf r})$ and $\phi_0({\bf r})$ is important, the scattering
length in the presence of light can be found from the asymptotic form
of $\phi({\bf r})$ at large distances:
$\phi({\bf r})\rightarrow 1-a/r$ for $r\rightarrow\infty$.
At large $r$ the Green function
$G({\bf r},{\bf r}')=m\phi_0({\bf r}')/4\pi\hbar^2r$, and
Eq.(\ref{phires}) yields
\begin{equation}         \label{a}
a=\overline{a}+\frac{(\Omega^2/\Delta\varepsilon_{\nu})\tilde\beta}
{\delta_{\nu}+(\Omega^2/\Delta\varepsilon_{\nu})\beta+i\Gamma}
r_t,
\end{equation}
with $\overline{a}$ the scattering length in the absence of light, and
the numerical factor $\tilde\beta=(m\Delta\varepsilon_{\nu}
/4\pi\hbar^2 r_t)|I_0|^2$.
It should be emphasized that Eq.(\ref{a}) is valid for any ratio
between $|\delta_{\nu}|$ and
$(\Omega^2/\Delta\varepsilon_{\nu})\beta$.

Unless $\varepsilon _{\nu}$ and $\left| \delta\right| $ are huge, the
turning point separation $r_t$ is large enough for $\phi _0$ and
$\tilde\phi_0$ to be smooth functions of $r$ at distances $r\sim r_t$
where the main contribution originates to the integrals in the
equations for $\beta$, $I_0$ and $\tilde\beta$.
Putting $\phi _0(r)=\phi _0(r_t)$,
$\tilde\phi_0(r)=\tilde\phi_0(r_t)$ in the integrands of these equations
and using a linear approximation for $V(r)=-2d^2/r^3$ in the vicinity
of $r_t$, we obtain
\begin{equation}            \label{gammasmooth}
\tilde\beta=0.8\pi^2 \phi_0^2(r_t);\,\,\,\,\,
\beta=0.8\pi^2 f_0(r_t)\phi_0(r_t).
\end{equation}
The function $f_0(r)=r\tilde\phi_0(r)$ is tending to $1$
for $r\rightarrow\infty$.
For the level spacing the WKB approximation gives
\begin{equation}       \label{deltae}
\Delta\varepsilon_{\nu}=1.9\pi\varepsilon_{\nu}(r_t/r_0)^{1/2}
\ll\varepsilon_{\nu}.
\end{equation}
The characteristic distance $r_0=md^2/\hbar^2$.
For alkali atoms $r_0$ greatly exceeds the optical wavelength
(\mbox{$r_0\agt 10^5$ \AA}) and, hence, $r_0\gg \lambdabar\gg r_t$.

The presence of other bound $p$ levels and continuum states of the
excited quasimolecule changes Eq.(\ref{Gex}) for the Green function
$\tilde G$.
Our analysis, relying on the exact expression for $\tilde G$, shows
that in order to omit the contribution of virtual transitions to these
states and, hence, retain the validity of Eq.(\ref{a}) it is
sufficient to have $|\delta_{\nu}|$ and $\Omega$ much smaller than
the level spacing $\Delta\varepsilon_{\nu}$.
The condition $\Omega\ll\Delta\varepsilon_{\nu}$ leads to
important physical consequences.
The radiative transitions occur in a narrow range of distances
near $r_t$, characterized by the width
$\Delta r\sim r_t(r_t/r_0)^{1/3}$.
As the characteristic velocity in this region $v\sim
\sqrt{\varepsilon_{\nu}\Delta r/mr_t}$, the interaction time
$\Delta t\sim\Delta r/v$ of the quasimolecule
with light is such that $\Omega\Delta t\!\sim \!
(\Omega/\Delta\varepsilon_{\nu}\!)(\!r_t/r_0\!)^{1/6}\!\!\ll \!1$.
Then, turning to a classical picture, one can say that
the ``population'' of the excited quasimolecular state will be
small.
This ensures the absence of effects
analogous to power broadening in the single atom case.

The light changes the real part of the scattering length
and introduces an imaginary part.
The frequency dependence of ${\rm Re}a$ and ${\rm Im}a$
has a resonance structure:
\begin{eqnarray}             \label{ares}
\!{\rm Re}a\!=\!\overline{a}\!+\!\frac{\Omega^2\tilde\beta
\zeta_{\nu}}{\Delta\varepsilon_{\nu}(\zeta_{\nu}^2\!+
\!\Gamma^2)}r_t;\,\,
{\rm Im}a\!=\!-\frac{\Omega^2\tilde\beta\Gamma}
{\Delta\varepsilon_{\nu}(\zeta_{\nu}^2\!+\!\Gamma^2)}r_t,\!\!\!
\end{eqnarray}
where $\zeta_{\nu}=\delta_{\nu}+(\Omega^2/\Delta\varepsilon_{\nu}
)\beta$.
The real part determines the mean field interaction between atoms.
The light-induced change of this interaction is given by
\begin{equation}      \label{taua}
n(\tilde U-\overline{\tilde U})\equiv\hbar \tau_a^{-1}=
4\pi\hbar^2({\rm Re}a-\overline{a})n/m.
\end{equation}
The imaginary part of $a$ originates from the photoassociation process
in pair collisions, followed by spontaneous emission.
The inverse decay time due to this process is
\begin{equation}               \label{pa}
\tau_{pa}^{-1}=8\pi\hbar|{\rm Im}a|n/{m}.
\end{equation}
Exactly at resonance ($\zeta_{\nu}=0$) the mean field interaction is the
same as in the absence of light, and the photoassociation rate is the
largest.

For small Rabi frequency  Eq.(\ref{ares}) goes over into the
result of perturbation theory and both $\tau_a^{-1}$ and $\tau_{pa}^{-1}$
are proportional to $\Omega^2$.
The former can be treated as a ``light shift'' of
the mean field interaction and the latter will be nothing else than
the ordinary photoassociation rate at a low light power.
For $(\Omega^2/\Delta\varepsilon_{\nu})\beta\gg |\delta_{\nu}|$
the driving light shifts the interacting pair of atoms out of
resonance.
As the corresponding shift is proportional to $\Omega^2$, the
light-induced change of the mean field interaction becomes
independent of $\Omega$. It will be determined by Eq.(\ref{taua}) with
${\rm Re}a-\overline{a}=(\tilde\beta/\beta)r_t$. On the contrary,
the photoassociation rate (${\rm Im}a$) decreases
as $1/\Omega^2$.

The amplitude of binary interaction, affected by light, undergoes
damped oscillations and reaches its stationary value
(\ref{a}) on a time scale of order $\Gamma^{-1}$ (for
\mbox{$\Gamma\ll max\{|\delta_{\nu}|,\Omega\}$} it averages
to Eq.(\ref{a}) much faster).
This is much
shorter than the characteristic response time of a dilute trapped
gas, which cannot be faster than $\tau_a$.
To have an appreciable modification of the mean field interaction
without excessive photoassociation, $\tau_a$
should be short compared to $\tau_{pa}$, i.e., the condition
\begin{equation}       \label{RI}
|{\rm Re}a|\gg |{\rm Im}a|
\end{equation}
should hold. As follows from Eq.(\ref{ares}), this is the case for
$|\zeta_{\nu} |\gg\Gamma $. The change of
the scattering length \mbox{${\rm Re}a-\overline{a}\approx(\Omega^2
\tilde\beta/\Delta\varepsilon_{\nu}\zeta_{\nu})r_t$} can
exceed $r_t$, whereas the imaginary part of $a$ will be much smaller.
The scattering length can be changed in both directions simply by
changing the sign of $\zeta_{\nu}$.

In addition, the time $\tau_a$ should be much smaller than
the decay time $\tau_r$ due to the photon recoil of single atoms,
caused by light scattering. Since $\tau_r^{-1}=(\Omega/\delta)^2\Gamma/2$,
this is the case for
($|\delta|\approx\varepsilon_{\nu}$)
\begin{equation}            \label{ineq}
n\lambdabar^3\gg |\zeta_{\nu}|/4\Delta\varepsilon_{\nu},
\end{equation}
as follows from Eqs. (\ref{ares}), (\ref{gammasmooth}) and
(\ref{deltae}) assuming
$|\zeta_{\nu}|\gg \Gamma$ to simultaneously satisfy condition
(\ref{RI}).
As $|\zeta_{\nu}|\ll \Delta\varepsilon_{\nu}$, the inequality
(\ref{ineq})
is not in contradiction with our starting assumption (\ref{n}) and
can be fulfilled in alkali atom gases at densities
$n\sim 10^{13} - 10^{14}$ cm$^{-3}$ by an appropriate
choice of $\Omega$, the level $\nu$ and $\delta_{\nu}$.

All the above results remain valid for finite momenta of colliding
atoms, $k\ll {\rm min}\{r_t^{-1},|\overline{a}|^{-1},|a|^{-1}\}$.

We performed calculations for $^7$Li by using spectroscopic
information on the location of bound $p$ levels in the excited
electronic state $^3\Sigma_g^{+}$ \cite{Li}. The potential of
interaction in the ground state $^3\Sigma_u^{+}$ was taken
from \cite{Pot}, the scattering length in the absence of light being
\mbox{$\overline{a}\approx -14$ \AA}. Eq.(\ref{ares}) was used to
calculate the scattering length $a$ under the influence of light nearly
resonant for vibrational $p$ levels of the $^3\Sigma_g^{+}$ state, with
quantum numbers ranging from $\nu=77$ ($\varepsilon_{\nu}=2.8$K) to
$\nu=66$ ($\varepsilon_{\nu}=28.7$K).
We find that for $\Omega$ in the range $5-
40$ mK (light power ranging from  $10$ to $1000$ W/cm$^2$) it is
possible to significantly change the scattering length and even make
it positive while maintaining $|{\rm Im}a|\ll |{\rm Re}a|$ (see Fig.1).
The recoil loss time $\tau_r$ varies from $100$ to $1$ ms.

Our results show the feasibility to optically manipulate the mean field
interaction between atoms and open prospects for new optical experiments in
trapped gases. For example, once a gas is in a Bose-condensed state,
instantaneous switching of the sign of $a$ changes the sign of the
non-linear interaction term in the Ginzburg-Gross-Pitaevskii equation for
the condensate wavefunction and causes the trapped condensate to evolve in a
completely different way than a condensate set into motion by changing the
trap frequency. The evolution will involve two time scales: $\tau_a$ and
the inverse trap frequency $\omega _t^{-1}$, and continue after the light is
switched off. Because of the light-induced decay processes, the light should
be switched on only for a time much shorter than $\tau_r$. Hence,
besides the above discussed condition $\tau_a\ll \tau_r$,
experiments should be arranged such that $\omega _t\tau_r\gg 1$. This is
feasible with the above values for $\tau_r$. As in most cases
$\tau _r$ will be much smaller than the characteristic time for elastic
collisions, the evolving condensate will not be in equilibrium
with above-condensate particles.

In trapped gases with negative scattering length one may expect a
stabilization of the condensate by switching $a$ to positive values.
Of particular interest is the case where the sign of $a$ is
switched from positive to negative. In a quasihomogeneous
Bose-condensed gas ($n\tilde U\gg \hbar\omega_t$) this should induce
a ``collapse'' of the condensate, caused by elastic interatomic
interaction.
The investigation of this phenomenon is of fundamental interest.

We acknowledge fruitful discussions with M.W. Reynolds and T.W. Hijmans.
This work was
supported by the Dutch Foundation FOM, by NWO through project
NWO-047-003.036, by INTAS and by the Russian Foundation for Basic Studies.%

\begin{figure}[tbp]
\caption{The scattering length for $^7$Li as a function of
the frequency detuning of light, $\delta_{\nu}/\Gamma$,
from the excited bound $p$ level $\nu$: a) $\Omega=10$ mK,
$\varepsilon_{\nu}=9.1$K;
b) $\Omega=40$ mK, $\varepsilon_{\nu}=20.1$K.
The solid curve represents the real part of $a$, and the
dashed curve the imaginary part. The dotted line corresponds to the
scattering length in the absence of light.}
\label{1}
\end{figure}

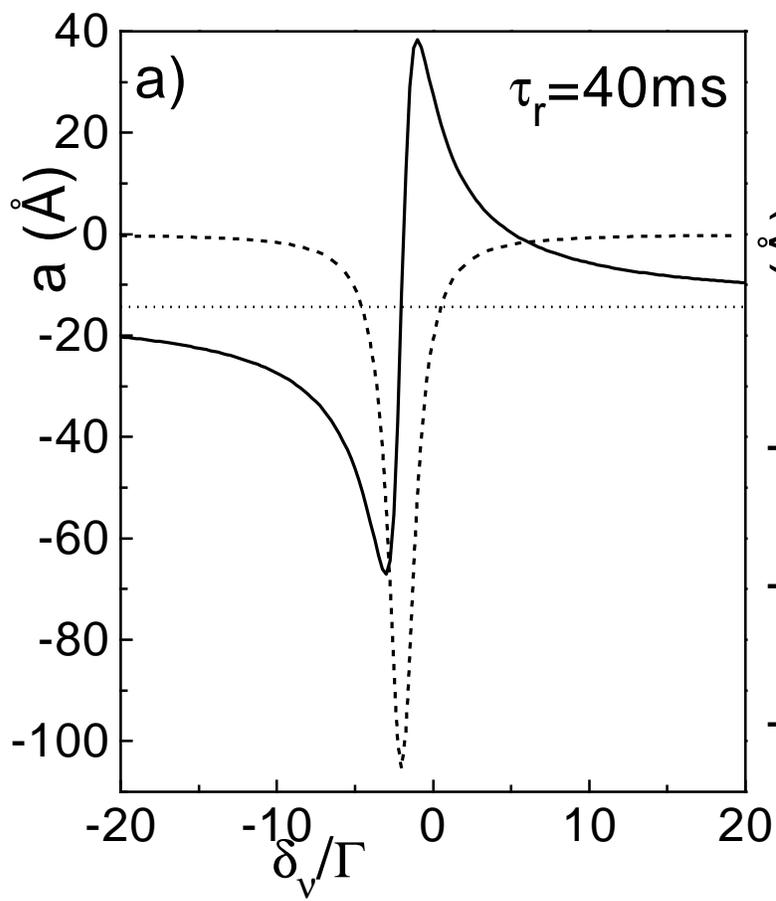
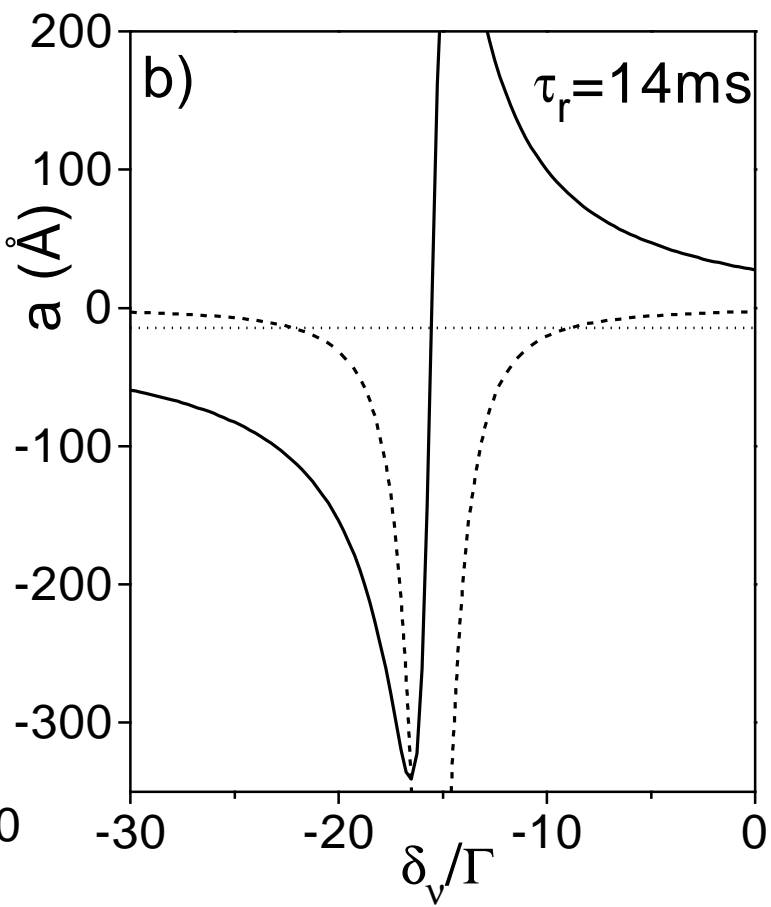
\end{document}